# Maximum Gene-Support Tree

Yunfeng Shan[1,2] and Xiu-Qing Li[1]

[1]Molecular Genetics Laboratory, Potato Research Centre, Agriculture and Agri-Food Canada, 850 Lincoln Rd, P.O. Box 20280, Fredericton, New Brunswick, E3B 4Z7, Canada. [2]Department of Natural History, Royal Ontario Museum, Toronto, Ontario M5S 2C6, Canada.

**Abstract:** Genomes and genes diversify during evolution; however, it is unclear to what extent genes still retain the relationship among species. Model species for molecular phylogenetic studies include yeasts and viruses whose genomes were sequenced as well as plants that have the fossil-supported true phylogenetic trees available. In this study, we generated single gene trees of seven yeast species as well as single gene trees of nine baculovirus species using all the orthologous genes among the species compared. Homologous genes among seven known plants were used for validation of the finding. Four algorithms—maximum parsimony (MP), minimum evolution (ME), maximum likelihood (ML), and neighbor-joining (NJ)—were used. Trees were reconstructed before and after weighting the DNA and protein sequence lengths among genes. Rarely a gene can always generate the "true tree" by all the four algorithms. However, the most frequent gene tree, termed "maximum gene-support tree" (MGS tree, or WMGS tree for the weighted one), in yeasts, baculoviruses, or plants was consistently found to be the "true tree" among the species. The results provide insights into the overall degree of divergence of orthologous genes of the genomes analyzed and suggest the following: 1) The true tree relationship among the species studied is still maintained by the largest group of orthologous genes; 2) There are usually more orthologous genes with higher similarities between genetically closer species than between genetically more distant ones; and 3) The maximum gene-support tree reflects the phylogenetic relationship among species in comparison.

## Introduction

Living organisms survive their environment through genetic variations such as transposition (McClintock, 1984), gene conversion (Archibald and Roger, 2002), horizontal gene transfer (Doolittle, 1999), adaptive selection (Logares et al. 2007), mutation or recombination (Vuli et al. 1999). This increasing genetic divergence of species makes it a challenge to reconstruct the true trees and to evaluate to what degrees the genes still retain their species relationship in taxa.

Various taxonomic groups such as some known plants have a well corroborated phylogeny or true tree that is based on combined support from fossil records and morphological characteristics (Russo et al. 1996). Such a set of organisms provides a reference for evaluating the reliability of molecular data-based alternative methods for determining phylogenetic relationships. Historically, determining the phylogeny of microbes was difficult due to the lack of discernible morphological characters (Fitz-Gibbon and House, 1999). Molecular phylogenetics has made great progress in studying the evolutionary relations among taxa, although incongruence in the phylogenetic tree reconstruction occurs from the methods used and genes studied (Russo et al. 1996; Doolittle, 1999; Baldauf et al. 2000; Rokas et al. 2003; Philippe et al. 2005; Simpson et al. 2006). Recently, whole genome sequences of a number of species became available, and it has increased the possibility to reconstruct a true tree through genome scale phylogeny (Rokas et al. 2003; Philippe et al. 2005).

There are mainly two alternative approaches for reconstructing genome-scale phylogenetic trees. The first is to concatenate many sequences head-to-tail into one and then reconstruct a tree (Kluge, 1989; Huelsenbeck et al. 1996; Yang, 1996; Rokas et al. 2003; Soltis et al. 2004). The second approach is to reconstruct many single-gene trees and then use the resulting trees to infer a majority rule consensus tree (Herniou et al. 2001; Gadagkar et al. 2005).

Yeasts and viruses are two important groups of model organisms for studying evolution and phylogenetics. The most accepted tree for representing the true tree of the seven yeast species was from the phylogenetic analysis of the concatenated sequence of 106 orthologous genes (Rokas et al. 2003). Similarly, the "true tree" of nine baculoviruses has been established from 63 shared gene sequences (Herniou et al. 2001).

Although fossil-based and molecular data-based phylogenetic analyses have been documented in various organisms, it is unknown to what extent the orthologous genes are divergent in term of tracing relations among species. In this study, gene by gene phylogenetic analysis of yeasts, baculoviruses, and plants confirmed that the most frequent gene tree among species compared is actually the true tree. The method, or called "maximum gene-support tree" approach may provide a potential tree reconstruction method that overcome incongruence in molecular phylogenies.

## Methods and Datasets

### Source of sequence data sets

Three data sets were utilized. The first data set contained 106 gene sequences from seven yeast species. These sequences have been previously analyzed using the genome-scale approach of concatenated alignment (Rokas et al. 2003). The yeast data set was retrieved from the *Saccharomyces* genomes database (http://www.yeastgenome.org). *S. bayanus*, *S. castellii*, *S. cerevisiae*, *S. kluyveri*, *S. kudriavzevii*, *S. mikatae*, and *S. paradoxus* were included. The fungus *Candida albicans* was included as the outgroup species. The second data set included 63 shared gene sequences from nine completed baculovirus genomes, as described by Herniou et al. (2001). The third data set contained 36 common homologous gene sequences from seven higher green plants, established by BLASTN (v2.2.6) search with the highest BLASTN score hit (e-value <0.0009) against NCBI nr/nt database using available *Ginkgo biloba* genes one by one. These sequences were retrieved from GenBank. The plant species included two gymnosperms, *Picea glauca* and *Pinus taeda;* two monocots, *Oryza sativa* and *Triticum aestivum;* and two dicots, *Populus tremula* and *Arabidopsis thaliana*. *Ginkgo biloba* was specified as the outgroup species. These species were selected because their phylogeny is well corroborated by the fossil record and morphological characters (Cronquist, 1981; Panchen, 1992; Campbell, 1993).

### Phylogenetic analysis

For the yeast and plant data sets, individual gene sequences were aligned using ClustalX with default settings (Thompson et al. 1997). All gene alignments were manually edited to exclude insertions or deletions and uncertain positions from further analysis. The phylogenetic analysis software PAUP* (Version 4.0b10) (Swofford, 2002) was used for tree inference based on four methods: MP, ME, NJ, and ML. Each nucleotide data set was analyzed under the optimality criteria of maximum parsimony for MP, distance for ME and NJ, and maximum likelihood for ML. The MP analyses were performed with unweighted parsimony. The ME, NJ and ML analyses were performed assuming the HKY85 model of nucleotide substitution. For the NJ analysis on amino acids, the absolute difference was used. The bootstrap consensus tree was searched using the branch-and-bound algorithm for MP and ML on nucleotides, and the full heuristic search was used for ME and NJ based on a 50% majority rule. 1000 replicates were used for all tests except for the ML, where 100 replicates were completed. Random sampling of genes was performed using a random number generator. For the baculovirus data set, only the phylogenetic trees obtained by Herniou et al. (2001) with the MP method were used.

### The maximum gene-support tree approach

From the yeast data set, bootstrap consensus trees were recovered using all 106 individual genes with seven combinations of four methods (ME, ML, MP and NJ) for nucleotides or three methods (ME, MP, NJ) for deduced amino acids. Tree distances for all pairwise comparisons among trees were calculated using the symmetric difference metric by PAUP* (Swofford, 2002) and PHYLIP (Felsenstein, 1989). This is the number of steps required to convert between two trees, that is, the number of branches that differ between a pair of trees (Robinson and Foulds, 1981). Two trees with identical topology have a tree distance of zero. For the baculoviruses, the comparison of topologies between the MP trees using the Shimoaira-Hasegawa (SH) test were directly cited from Herniou et al. (2001). For the plant data set, comparisons between the trees were performed manually.

The index of gene-support is the number of genes that support a certain topology. The resulting numbers of genes were calculated for all unique trees from the results of each method. A maximum gene-support tree was defined as a unique tree that was recovered by the highest number of genes among all the trees generated. The statistics analyses were performed using the SAS system for Windows V8.

### Re-sampling for subsets of genes

Subsets of genes were randomly re-sampled using a random number generator. Ten replicates were used for each initial number of re-sampled genes. Precision was defined as the percentage of the number of congruent trees divided by the total number of trees. 100% precision was used as a criterion to determine the minimum number of genes required to overcome incongruence.

An executable program in C language for calculating frequencies of unique trees from tree distance data is available from the authors upon request (shan@cs.dal.ca; lixq@agr.gc.ca).

## Results

### Incongruence among different individual-gene phylogenies

Wide incongruence was observed among individual gene trees. The 106 individual genes inferred 20 to 51 unique trees for the 7 yeast species using 7 combinations of 4 phylogenetic methods with nucleotides or 3 methods with deduced amino acids (Table 1). Nucleotides inferred fewer unique trees (20–38) than amino acids (40–51) (Table 1). For example, the occurrence of the maximum gene-support tree for MP based on nucleotides was 37, while that based on amino acids was only 14 of 106 genes. Trees recovered from amino acids had more incongruence and less gene-support than those from nucleotides.

### The maximum gene-support tree

The maximum gene-support trees for the seven yeast species from different methods (MP, ME, ML, NJ) based on both nucleotide sequences and amino acid sequences were all identical (Fig. 1). The maximum gene-supports of the unique trees recovered by 106 genes were 37, 33, 25, 28 for MP, ME, NJ, and ML on nucleotides, respectively, and 14, 14, and 17 for MP, ME, and NJ on amino acids, respectively (Table 1). The maximum gene-support percentages were 35%, 31%, 24%, 26% for MP, ME, NJ, ML on nucleotides, respectively, and 13%, 14%, and 16% for MP, ME, and NJ on amino acids, respectively. The second most gene-support percentages were considerably smaller, 9%, 15%, 22%, 9% for MP, ME, NJ, ML on nucleotides, respectively, and 10%, 9%, and 8% for MP, ME, and NJ on amino acids, respectively (Table 1). Gene-support is defined as the number of genes that infer the same unique tree. The occurrence of maximum gene-support trees for nucleotides consistently had greater gene-support values than those for amino acids.

Table 1. Maximum gene-support (MGS), weighted maximum gene-support (WMGS), the second highest gene-support (2nd HGS), weighted second highest gene-support (2nd WHGS), number of unique trees (NUT), and threshold gene number (TGN) required to overcome incongruence based on a data set of 106 genes from seven yeast species*.

|             | MGS      | WMGS     | 2nd HGS  | 2nd WHGS | NUT | TGN |
|-------------|----------|----------|----------|----------|-----|-----|
| **Nucleotides** |      |          |          |          |     |     |
| MP          | 37(35%)  | 42(40%)  | 10(9%)   | 13(12%)  | 31  | 15  |
| ME          | 33(31%)  | 32(30%)  | 16(15%)  | 17(16%)  | 23  | 26  |
| NJ          | 25(24%)  | 25(23%)  | 23(22%)  | 23(22%)  | 20  | 106 |
| ML          | 28(26%)  | 34(32%)  | 9(9%)    | 12(11%)  | 38  | 25  |
| **Amino acids** |      |          |          |          |     |     |
| MP          | 14(13%)  | 18(17%)  | 11(10%)  | 9(9%)    | 51  | 55  |
| ME          | 14(14%)  | 14(14%)  | 10(9%)   | 10(9%)   | 40  | 50  |
| NJ          | 17(16%)  | 17(16%)  | 8(8%)    | 10(9%)   | 40  | 50  |

*Gene-support: number of genes that infer a unique tree; Gene-support percentage in parenthesis: the percentage of a gene-support divided by total genes; Number of unique trees: number of unique trees inferred from 106 genes; Threshold gene number: the minimum number of genes required for overcoming incongruence.

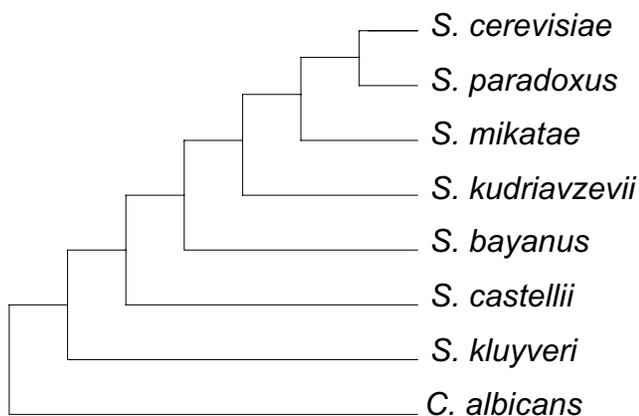

**Figure 1.** The rooted tree with the maximum gene-support inferred from 106 genes of seven yeast species. The outgroup in the analysis was *C. albicans*. The single gene trees were recovered using bootstrap consensus with a 50% majority rule.

## Gene length and tree distance

A significant negative correlation between gene length and symmetric distance of a tree from the maximum gene-support tree was observed (Fig. 3). The greater the gene length, the shorter symmetric distance the tree was to the maximum gene-support tree.

## The weighted maximum gene-support tree

Because sequence length is an important factor affecting single gene tree inference, adjustments were conducted by means of a weight factor, which is equal to the gene actual length divided by the average length of all the genes. The average sequence length of 106 genes was 1198 bps. The weight factors of the 106 genes were distributed between 0.33 and 2.50 (Fig. 2). For example, if the weight factor of gene A is 0.33, the value it contributes to the weighted maximum gene support is 0.33. No evident differences between the weighted and the unweighted maximum gene-supports were observed in any of the seven combinations in this study (Table 1). The weighted maximum gene-support tree was also consistent with the maximum gene-support tree (Fig. 1).

## Gene-support and tree distance

There was a significant correlation between gene-support and symmetric distance of a tree from the maximum gene-support tree (Fig. 4). The greater the gene-support for a tree, the closer the tree was to the maximum gene-support tree. The topologies of the second gene-support trees were very similar to the maximum gene-support tree. Generally, only one or two steps were required to convert between the two trees.

## The minimum number of genes required to overcome incongruence

The precision of MP trees based on nucleotide sequences inferred from 5, 10, 15 or 20 genes was 80%, 90%, 100% or 100%, respectively. Therefore, at least 15 genes were required to overcome incongruence for the seven yeast species studied. For other methods, the minimum number of genes was 26, 106, 25 for ME, NJ, ML, respectively, based on nucleotides and 55, 50, and 50 for MP, ME, and NJ, respectively, based on amino acids (Table 1). Rokas et al. (2003) found that the number of genes sufficient to support all branches of the species tree was 20 based on the concatenated alignments of 106 genes from the same seven yeast species. The number varied with methods and taxa.

The minimum size (number of genes) in the dataset required for generating a MGS tree generally decreased with increased maximum gene-support percentages (Table 1). This number depended not only on the maximum gene-support, but also on the second highest gene-support. The closer the two values were, the more difficult it was to identify the congruent tree. This is illustrated by the NJ method using nucleotides, where the maximum gene-support was 25 and the second highest gene-support was 23 (Table 1). In this case, the minimum required number of genes was 106 because the two trees were very similar and differed by only one branch. In contrast, for the NJ method on amino acids, the minimum number of genes was only 30 when the maximum gene-support was 18 and the second highest gene-support percentage was 8.

The maximum gene-support, the second highest gene-support, and the gap between them expanded when more genes were involved although the maximum gene-support percentages and the second highest gene-support percentages did not increase (Table 2). At the same time, precision increased. Therefore, higher confidence is obtained when more genes are involved.

## Validation using data sets of other taxa

Using 63 shared genes from nine complete baculovirus genomes, the maximum gene-support based on MP with amino acids was 7. The maximum gene-support tree was identical to the tree recovered from the concatenated alignments, but not from the majority rule approach (Herniou et al. 2001). The maximum gene-support trees (Fig. 5) of 36 shared single genes (Table S1) of seven green plants using three methods were the same trees recovered by fossil record and morphological characters (Cronquist, 1981; Panchen, 1992; Campbell, 1993). The maximum gene-supports were 8, 8, and 6 for 36 genes using MP, ME and NJ, respectively.

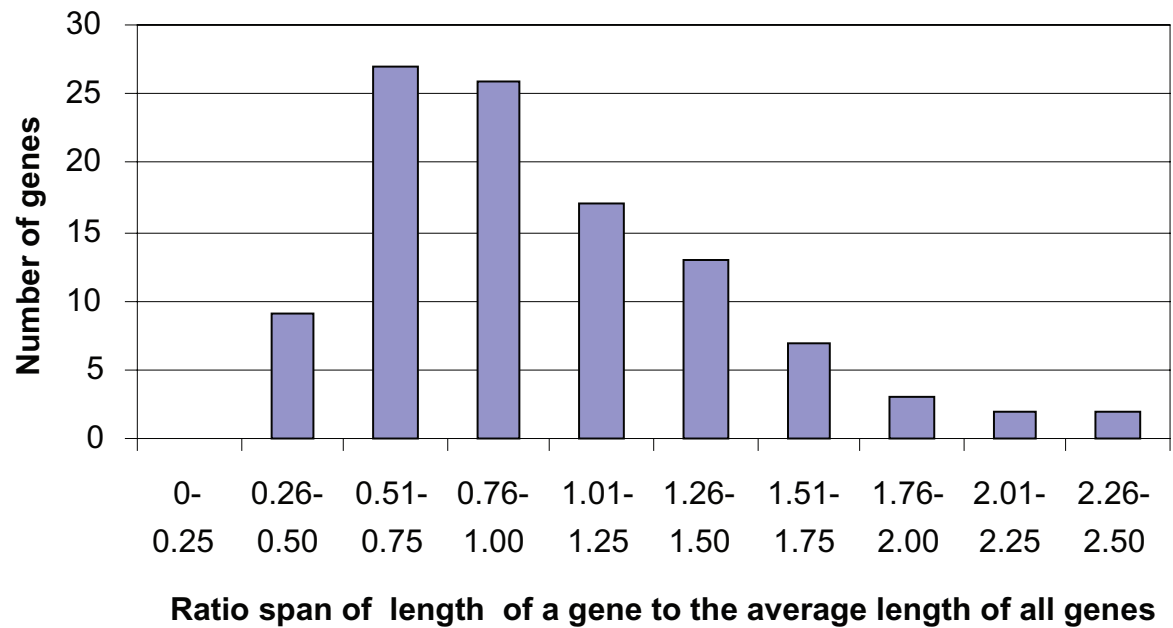

**Figure 2.** Distribution of sequence length of 106 genes.

## Probability of a gene suitable for reconstructing a true tree

It is a usual practice to use more than one phylogenetic method to analyze the same data set in order to test congruence between trees. In Table 1, the maximum gene-support tree reconstructed by MP on nucleotides for the seven yeast species was 37, which means that any of these 37 genes will reconstruct this tree. Similarly, ME identified 33 genes that meet this requirement (Table 1). However, only 14 genes were present in the maximum gene-support trees generated by both the MP and ME methods on amino acids. When this analysis was extended to all seven calculations (four algorithms using nucleotide sequences and three algorithms using amino acids), only 1 of these 106 genes, YDR176W of 711 nucleotide bp, was represented in all maximum gene-support trees. The results suggest that selecting a gene suitable for reconstructing a tree that represents the congruent phylogenetic relationship among taxa with several methods is very difficult. These genes cannot be easily identified, since they appear to be quite rare.

## Discussion

This study uses orthologous/homologous genes to demonstrate that the gene tree supported by the largest group of genes is identical to the true tree for the seven plant species tested or the best genome-based phylogenetic tree. The maximum gene-support tree provides an evaluation of the degree of evolution of genes at the genome level. Further studies using more living organisms can verify this intriguing evolutionary phenomenon.

This study also demonstrates that incongruence in molecular phylogenies can be caused by both genes and methods. The maximum gene-support tree approach is different from the approaches used by the majority rule consensus tree or the concatenated alignments tree. Both the majority rule consensus tree and the maximum gene-support tree approaches are based on a majority rule method. The first step for the two approaches reconstructs

Shan and Li

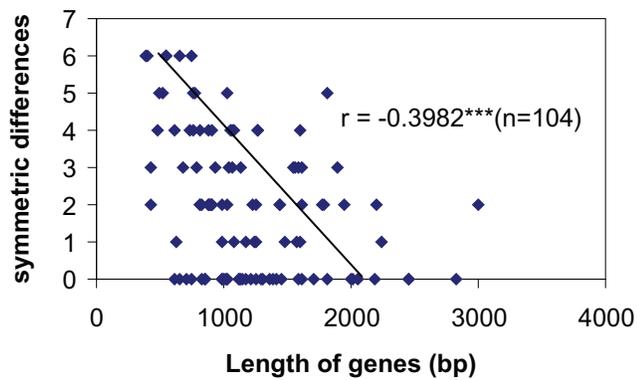

**Figure 3.** The correlation between symmetric differences from trees to the MSG tree and gene lengths.

all single gene trees. However, the subsequent steps are different. The majority rule consensus tree method counts the occurrences of each subtree, creates nodes for the subtrees that occur in a majority of input trees, i.e. the *majority subtrees,* and "hooks them" together into a tree (Gadagkar et al. 2005). For the concatenated alignments approach, the first step is to concatenate small alignments into one large alignment, and then a tree is reconstructed using the large alignment (Kluge, 1989; Huelsenbeck et al. 1996; Yang, 1996; Rokas et al. 2003; Soltis et al. 2004). The concatenated alignments approach apparently reconstructs more accurate trees than the majority rule consensus tree approach (Gadagkar et al. 2005). In contrast, the maximum gene-support tree approach directly compares whole trees, counts the occurrences of unique trees, and finds the tree that is supported by the greatest number of genes. The main advantage of this approach is its simple algorithm. A maximum gene-support tree is a representation of a consensus tree from multiple trees inferred from individual genes. The majority rule consensus tree is useful only if topologically is identical with one of the original trees (Wiley et al. 1991), while the maximum gene-support tree always meets this requirement. An absolute majority rule is not suitable for this approach because too many candidate trees are produced using different methods. For example, in this study 20 to 51 unique trees were inferred from 106 yeast genes using seven combinations of sequence types and methods. We refer to this relative majority rule tree as a maximum gene-support tree and a maximum frequency gene tree in order to avoid confusion with a majority rule consensus tree or a bootstrap tree with a majority rule (Felsenstein, 1989; Gadagkar et al. 2005).

The maximum gene-support tree approach showed advantages over the concatenated alignment approach when the seven combinations of methods and sequence types were tested. The trees were reconstructed by ME, NJ on amino acids and NJ on nucleotides using the concatenated alignments of yeasts (Fig. 6) because they were not reported by Rokas et al. (2003) and Phillips et al. (2004). The topologies of the trees were identical with those of MP and ML from the same concatenated alignments (Rokas et al. 2003), although 58%, 76% and 96% supports were observed on NJ trees (Fig. 6). On the other hand, the tree resulting from ME on nucleotides using concatenated alignments (Phillips et al. 2004) was different from the tree of MP and ML recovered by the same concatenated data set when base biases were not adjusted (Rokas et al. 2003). The concatenation of genes that share some biases can produce the incorrect phylogeny with strong support (Rokas et al. 2003) due to accumulation of systematic errors of base biases (Phillips et al. 2004). When gene sequences are concatenated, the deviated gene may over-contribute to the computing, and dominate or

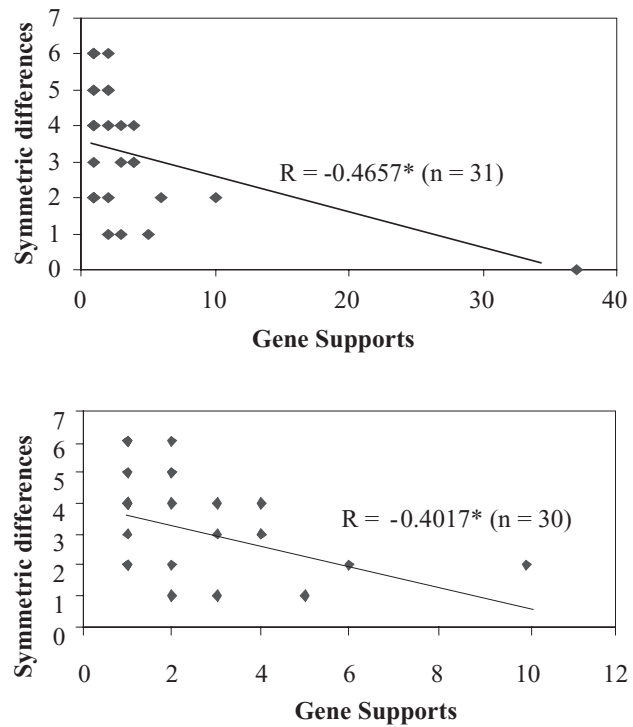

**Figure 4.** Relationship between gene-support percentage of unique trees and symmetric distances of the trees from the maximum gene-support tree. The symmetric difference is the number of steps required to convert between two trees. MP trees inferred from nucleotides were used here (no data shown for other methods included).
*Indicates statistical significance at the p = 0.05 level. Top panel: Full dataset; Bottom panel: After taking off the last point.

Table 2. The number of sampled genes, the maximum gene-supports (MGS), the second highest gene-supports (2nd_HGS), the differences between MGS and 2nd_HGS (DMGS), the maximum gene-support percentages (MGSP), the second highest gene-support percentages (2nd_HGSP), the differences of MGSP and 2nd_HGSP (DMGSP), and precisions*.

| Genes | MGS | 2nd_HGS | DMGS | MGSP % | 2nd HGSP % | DMGSP % | Precision % |
|---|---|---|---|---|---|---|---|
| 5 | 1.6(0.5) | 1.0(0) | 0.6(0.5) | 32.0(11.0) | 20.0(0) | 12.0(11.0) | 60 |
| 10 | 3.2(1.4) | 1.4(0.5) | 1.8(1.3) | 32.0(14.0) | 14.0(5.2) | 18.0(13.2) | 80 |
| 15 | 4.0(1.6) | 2.4(0.7) | 1.6(2.0) | 26.7(10.4) | 16.0(4.7) | 10.7(13.0) | 60 |
| 20 | 5.3(2.3) | 2.5(0.5) | 2.8(2.6) | 26.5(11.6) | 12.5(2.6) | 14.0(13.0) | 90 |
| 24 | 6.6(1.8) | 2.7(0.7) | 3.9(2.2) | 27.5(7.7) | 11.3(2.8) | 16.3(9.1) | 90 |
| 25 | 6.5(1.5) | 2.9(0.6) | 3.6(1.8) | 26.0(6.0) | 11.6(2.3) | 14.4(7.4) | 100 |
| 30 | 8.3(1.6) | 3.1(0.7) | 5.2(2.1) | 27.7(5.2) | 10.3(2.5) | 17.3(7.2) | 100 |
| 106 | 28 | 9 | 19 | 26.4 | 8.5 | 17.9 | 100 |
| r | 0.91*** | 0.86*** | 0.78*** | –0.11 | –0.44 | 0.07 | 0.55 |
| df | 64 | 64 | 64 | 64 | 64 | 64 | 6 |

*ML trees inferred from nucleotides (data not shown for other methods). Sample replicates: 10. Precision: the percentage of the number of congruent trees divided by the total number of trees.
Values in parenthesis are standard deviations of the values. ***: Significant correlation at $P \leq 0.001$ level. r: Correlation coefficient. Df: Degree of freedom.

sweep the signal (Doyle, 1992; De Queiroz, 1993; Miyamoto and Fitch, 1995). The maximum gene-support tree approach does not have this kind of problem or systematic error accumulation because each gene tree is separately reconstructed and contributes equally.

It is well known that the longer sequences of single genes usually tend to reconstruct better trees than shorter sequences. We showed that there is a significant negative correlation between gene length and symmetric distance of a tree from the maximum gene-support tree (Fig. 3). In order to remove the effects of gene length, adjustment was performed by average length of all sampled genes. In the datasets analyzed, weighted maximum gene-support (WMGS trees) did not show any difference from the maximum-gene support trees (MGS trees). It is unclear whether this just happened to the three datasets used or because each gene is an entity for certain functions despite the length difference. Since the sequence length effect is well known, the weighted maximum gene-support tree approach is recommended at this stage. Further research is required to determine whether the WMGS tree approach is biologically more sound than the MGS tree approach.

This maximum gene-support tree approach avoids repeating intensive computing of large data sets of genome-scale concatenated alignments. It took 19 days to complete the phylogenetic analysis using ML with concatenated alignments of just 36 genes of seven plant species on a PowerPC Macintosh computer with 1.2 GHz CPU. Other authors have previously commented on the computational limitations of the ML method using concatenated alignments (Wolf et al. 2002). The computation time for thousands of genes from higher eukaryotes would be even more unacceptable. In addition, when more gene sequences are involved, the concatenated approach requires all the computing processes to be repeated, while the maximum gene-support tree approach simply requires the addition of new single trees of the new genes. However, the ML method can still be an effective and efficient method with the maximum gene-support tree approach by distributing computing tasks to available PC computers since each tree is inferred by each gene independently. For the concatenation approach, a parallel version of ML is necessary, but this is not available in most laboratories.

When recovered trees include polytomies, a more logical approach would be to add all equally parsimonious trees recovered from a single gene to the total tree data set rather than first calculating consensus trees for each individual gene. One gene may contribute two or more trees for these genes while another gene contributes a single tree. Adjustment may be conducted by a contribution factor. The gene contribution may be divided by the number of equally parsimonious trees.

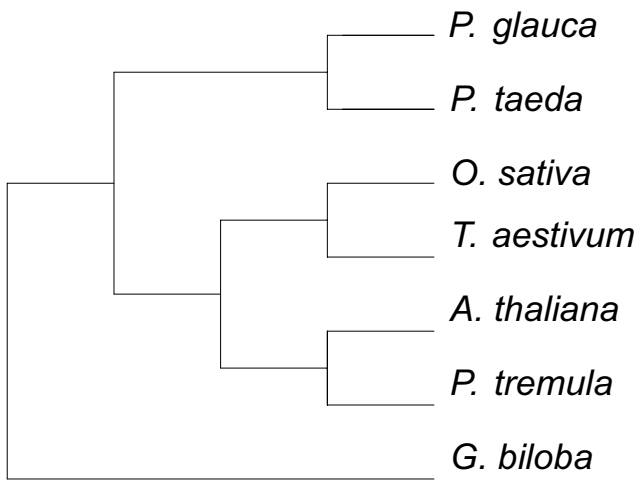

**Figure 5.** The rooted maximum gene-support tree based on 36 genes from seven plant species. *G. biloba* was specified as the outgroup.

When the gene number is small, the gene number for the maximum gene-support tree may be equal to that of the second-highest gene-support trees. As well, the gene support confidence can be very low, such as when only 2 genes return the same tree. For this situation, it is evident that the number of genes is too small to reach the minimum requirement for widely incongruent single gene trees. The solution is to involve more genes in the analysis (similar to increasing sample size in other investigations). As shown in Table 2, when only 5 genes were used, difference between maximum gene-support and the second highest gene-support was 0.6, thus the precision was 60%. The precision increased to 100% when 25 genes were used, and the gap between the maximum gene-support tree and the second highest gene-support tree was 3.6 (Table 2). The increased gap and gene-support enhance the confidence for reconstructing a phylogenetic tree. Evidently, gene-support percentages did not increase when more genes were included (Table 2). The jackknife method is suitable for re-sampling individual genes in order to determine the precision and to judge whether the required gene number threshold has been reached. If the maximum gene-support is very close to the second highest gene-support, it is difficult to identify the maximum gene-support tree, even though gene-support is rather large as shown by NJ on amino acids. One solution is still to include more genes. Since the tree distance between a maximum gene-support tree and a second highest gene-support tree differs by just one or two branches, cross-validation with other maximum gene-support trees inferred by other methods may be an alternative feasible approach.

Obtaining a sufficient number of shared genes may become difficult, and even unrealistic if too many taxa are involved. More orthologous genes are likely required when more species are tested. However, when many small trees are recovered using a minimum number of shared genes by means of the maximum gene-support tree approach, a larger picture of evolutionary relationships can gradually be reconstructed using a divide and conquer strategy of overlapping and connecting many smaller trees (Sanderson et al. 1998; Semple and Steel 2000).

As shown in Table 1, the gene-supports and its percentages of the maximum gene-support trees on nucleotides were greater than those on amino acids. When nucleotide sequences were used, more genes reconstruct the same tree, which means that nucleotide sequences may be more suitable for inferring species phylogeny. This result supports the hypothesis (Ayala et al. 1996) that evolution is more regular at the nucleotide level than at the protein level and, thus, more dependable as a molecular clock.

This maximum gene-support tree approach is likely an appropriate method to assess the phylogenetic relationship across certain range of taxa, as evident from the analysis of the three data sets (nine virus races, seven yeast species, a fungus, and seven botanically distant plants) in this study. The phylogenetic relationships have been previously identified using fossil records and morphological characteristics for these plants (Cronquist, 1981; Panchen, 1992; Campbell, 1993) is now with further congruent support from the maximum gene-support tree. More studies are still needed to establish the generality of using the maximum-gene-support tree model in phylogeny with a huge number of species when their genome sequences are available.

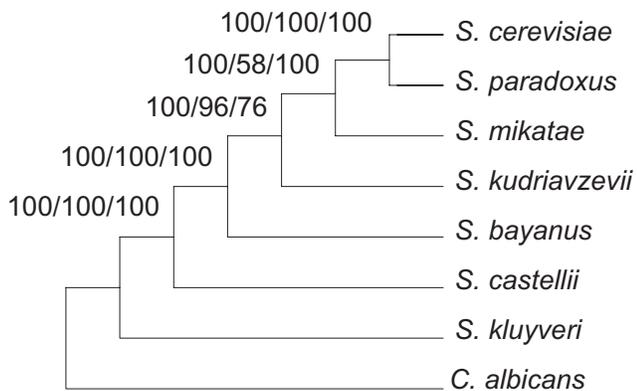

**Figure 6.** Phylogenetic analyses of the concatenated alignments of 106 genes from seven yeast species. Numbers above branches are bootstrap values (ME on amino acids/NJ on amino acids/NJ on nucleotides).

In a hypothetical scenario in which a million or more species are compared at the same time, a single gene's polymorphism, particularly for the short sequence genes, may not be useful in distinguishing all the species regardless of the degree of polymorphism the gene has in the population. In this scenario, it is unclear whether the maximum-gene-support tree is still a good representation of the true tree. To date, it is unlikely any of the phylogenetic methods are prefect, because each of them has their advantages and disadvantages. The maximum gene-support tree approach has its strength in comparing relatively close species because the approach is based on biological phenomenon, observed in the present study, that there are usually more orthologous genes with higher similarities between genetically closer species than between genetically more distant ones.

Two conclusions can be drawn from the present study: 1) The true tree relationship among species within each database studied is still maintained by the largest group of orthologous genes, although genes are of great divergence among organisms; and 2) The maximum gene-support tree, at least when the taxonomic range and the species number are not too large, is likely an effective novel approach for phylogenetic analysis with various advantages compared to existing approaches in the genome-scale or the large-gene-number-scale phylogenetic analysis.

## Acknowledgements

The authors are most grateful to Dr. Antonis Rokas, who kindly provided suggestions and the aligned yeast data sets and their trees. The authors also sincerely thank Dr. Richard Winterbottom and Dr. David De Koeyer for critical reading, comments, and useful discussion. The authors greatly appreciate helpful suggestions and comments of the two reviewers.

# Maximum Gene-Support Tree

Yunfeng Shan and Xiu-Qing Li

## Supplementary Material

**Table S1.** The names and GenBank GIs of the 36 genes of seven plant species.